\documentclass[twocolumn]{aa}
\usepackage{graphicx}
\usepackage{txfonts}
\usepackage{natbib}
\bibpunct{(}{)}{;}{a}{}{,}
\begin{document}
\title{INTEGRAL observations of the PSR~B1259-63~/~SS2883 system
 after the 2004 periastron passage}

\titlerunning{INTEGRAL Observations of \object{PSR B1259-63}}

\author
{S.E. Shaw\inst{1,2} 
\and M. Chernyakova\inst{2,4} 
\and J. Rodriguez\inst{3,2} 
\and R. Walter\inst{2,4} 
\and P. Kretschmar\inst{5,2} 
\and S. Mereghetti\inst{6}
}
\institute{School of Physics and Astronomy, University of
Southampton, SO17 1BJ, UK 
\and INTEGRAL Science Data Centre,
CH-1290 Versoix, Switzerland 
\and CEA Saclay, DSM/DAPNIA/SAp (CNRS FRE 2591), F-91191 Gif Sur
Yvette Cedex, France 
\and Observatoire de Gen\`eve, 51 Chemin des Maillettes,
CH-1290 Sauverny, Switzerland 
\and Max-Planck-Institut f\"ur
Extraterrestrische Physik,  87548 Garching, Germany 
\and Istituto di Astrofisica Spaziale e Fisica Cosmica, Sezione di Milano, via
Bassini 15, I-20133 Milano, Italy }

\offprints{\email{simon.shaw@obs.unige.ch}}

\date{Received now / Accepted then}

\abstract{The millisecond pulsar \object{PSR B1259-63} and the Be star \object{SS2883}
  are situated in a highly eccentric binary system, with orbital
  period $\sim$3.4~years.  We report on hard X-ray observations obtained with the \textit{INTEGRAL} satellite close to the
2004 periastron passage. These are the first imaging observations
of this system in the hard X-ray range ($>$20 keV) and allow the
emission of \object{PSR B1259-63} to be separated from that of a nearby
unidentified variable source which may have contaminated
previous hard X-ray observations of this system. Using the
\textit{IBIS/ISGRI} instrument we measured a flux of $\sim$2--6~mCrab 
in the 20--200~keV range, with a power law spectrum of photon index
$\Gamma = 1.3 \pm 0.5$.
\keywords{gamma-rays: observations -- pulsars:individual \object{PSR B1259-63} }
   }

\maketitle

%

\section{Introduction}
\label{sec:intro}
\object{PSR B1259-63} is a 47.6 ms radio pulsar in a highly eccentric
($e \sim 0.87$) orbit, with a period of 1236.7 days, around the
massive Be star \object{SS2883} \citep{1992MNRAS.255..401J}.  Be stars
are well known to be the
source of strong, anisotropic matter outflows via a diluted
polar wind and a denser equatorial disk
\citep{1988A&A...198..200W}.  Timing analysis of the system
indicates that the disk surrounding the Be star \object{SS2883} is highly
tilted and that the line of intersection of the disk and orbital
planes is almost perpendicular to the orbital major axis
\citep{1998MNRAS.298..997W}. Hence, the pulsar crosses the disk
twice, at approximately $\tau~\pm~20$ days, where $\tau$ denotes
the date of periastron, which is Modified Julian Date (MJD)
53071.4307 in this epoch.

Since its discovery, this source has
been intensively studied in the radio domain and, to a lesser
extent, at X-ray and gamma-ray energies.  Monitoring of the source
shows that unpulsed radio emission appears
after the first disk passage, at about $\tau-20$ days, and
persists until $\sim\tau+100$ days, with no significant pulsed
radio emission being observed around the periastron \citep[see][ and references therein]{2002MNRAS.336.1201C}.

The first X-ray detections of \object{PSR B1259-63} were made, near apastron of the
system, by the \textit{ROSAT} telescope in February 1992 -- February
1993 after earlier observations with \textit{Ginga} produced only
upper limits \citep{1994ApJ...427..978C,1995ApJ...441L..43G}.
Observations near the 1994 January 9 periastron, with the
 \textit{ASCA} satellite, showed that
the 1--10 keV luminosity  varied from $10^{34}$~erg~s$^{-1}$ at periastron to $10^{33}$~erg~s$^{-1}$ at apastron \citep{1995ApJ...453..424K}. At periods close to the disk crossings
($\tau-12$,~$\tau+17$ days) the 1--10 keV luminosity was
approximately twice higher than at the periastron. A reanalysis of
the \textit{ASCA} data found that the spectrum of the source was consistent with a
moderately absorbed  power law ($\rm{N}_{\rm{H}}=6\times10^{21}$~cm$^{-2}$)
with  photon index varying  from 1.96 at periastron, to 1.7 during
the time of the disk crossing, and to 1.6 at apastron
\citep{1999ApJ...521..718H}. No pulsed X-ray emission has been
detected from the system.

\begin{figure*}[t]
\centering
\includegraphics[width=17cm]{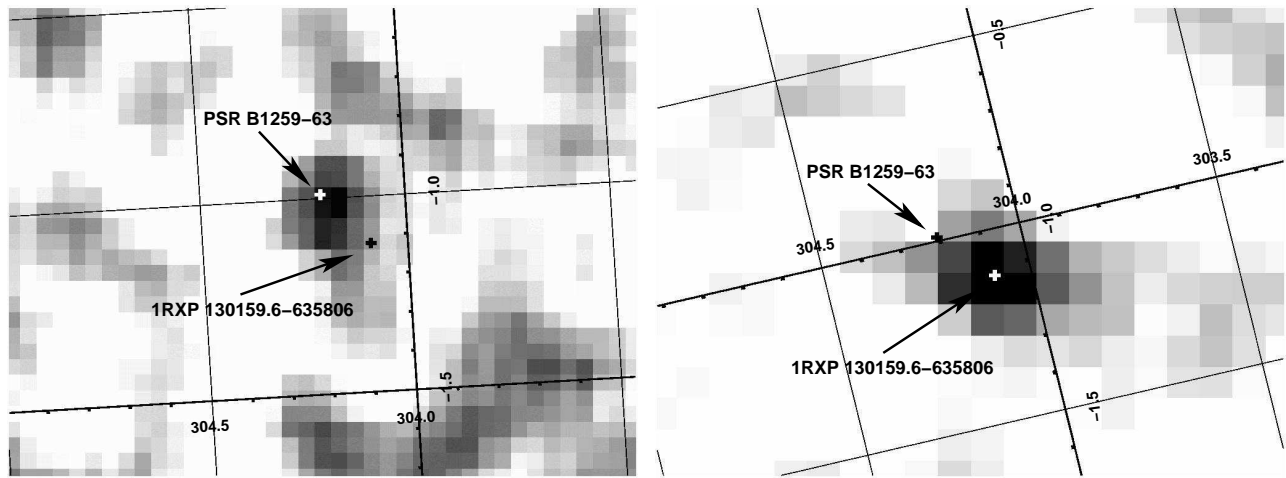}
 \caption{Mosaic images from the \textit{ISGRI} instrument in Galactic
   Coordinates.  Left:  30--50 keV image from the \textit{INTEGRAL}
   TOO observation conducted between 21--23 March 2004.  Emission from
   \object{PSR B1259-63} dominates, at a significance level of 4$\sigma$.
   Right: 20--60 keV image from the 7th February 2004 Galactic Plane Scan
   showing significant emission from
   1RXP~J130159.6-635806, at $\sim$13~mCrab.}
\label{fig:image}
\end{figure*}

Before the observations reported here, the only hard X-ray
measurement (40--300 keV) from \object{PSR B1259-63} was obtained by the
\textit{OSSE} instrument during monitoring of the 1994 periastron passage \citep{1995ApJ...447L.113G}.  Since \textit{OSSE} was a non-imaging detector, it cannot be excluded that other sources in this region
contributed to the measured flux. In fact, a variable source with a hard X-ray spectrum is present at only
10$'$ from \object{PSR B1259-63} (see Sect.~\ref{sec:integral}). Furthermore, the \textit{OSSE}
sensitivity was such that a significant detection could only be
obtained by summing all the data taken during a 20 day long
observation starting at $\sim \tau -6$ days.

Recently, the first detection of photons at very high energies ($>$
200 GeV) from this system has been obtained using  the
\textit{HESS} Cherenkov telescope. \citet{2004IAUC.8300....2B}
  report a flux corresponding to $\sim$~10\% of the
Crab Nebula from observations taken in the period between $\tau-10$ and
$\tau-2$ days.

\section{Data analysis and results}

\subsection{INTEGRAL}
\label{sec:integral}

Prompted by the \textit{HESS} high-energy detection, a Target
of Opportunity (ToO) observation was requested with the
\textit{INTEGRAL} satellite, which includes several imaging instruments
covering the $\sim$3 keV to
$\sim$10 MeV band \citep{2003A&A...411L...1W}.  The \textit{INTEGRAL}
ToO observation of \object{PSR B1259-63} was conducted
between 2004 March 21 12:34 -- March 24 12:13 UTC (from
$\tau~+14.1$ days to $\tau~+17.5$ days).  This time interval was
chosen based on the expected period for the second disk crossing
(20-25 March). 

We concentrate here mainly on data obtained with the \textit{IBIS} coded
mask telescope \citep{2003A&A...411L.131U} and in particular with its
lower energy detector \textit{ISGRI}
\citep{2003A&A...411L.141L}. \textit{ISGRI} operates in the 15 keV -- 1 MeV energy range, providing images with
an angular resolution of $\sim$12$'$ over a large field of view of
29$^{\circ}\times29^{\circ}$. Note that the source location
accuracy of coded mask imaging instruments is much better than the
nominal angular resolution.  For example, sources detected with a
significance of 10$\sigma$ can be located by \textit{ISGRI} with an
accuracy of 1$'$ \citep{2003A&A...411L.141L}. 

The \textit{ISGRI} data consist of $\sim$120 exposures, each lasting
1800~s and pointed according to a hexagonal dithering
pattern chosen to optimize the \textit{INTEGRAL} imaging performance.  Data were analysed using
a pre-release version of the \textit{ISDC OSA 4} software package, with images from individual
pointings being combined with the mosaic software described in
\citet{2004ApJ...607L..33B}.  The
target was not detected in any individual exposure. In
Fig.~\ref{fig:image} (left panel) we show a section of the
mosaic image obtained by summing the data in the 30-50 keV energy
range for the whole observation, which amounts to a total of 190 ksecs
of effective exposure time.
A source with  a statistical significance  of $\sim4\sigma$ is
detected at the position of \object{PSR B1259-63}.

This region of the sky was also briefly observed by \textit{INTEGRAL},
during the Core Program monitoring of the Galactic plane, on 2004
February 7 and 19. The \textit{ISGRI} image obtained on February 7,
corresponding to an integration time of only 8800~s,  is shown in
the right panel of Fig.~\ref{fig:image}. The source detected on
February 7  coincides with  the X-ray source \object{1RXP J130159.6-635806}.
This source, located approximately 10$'$ from \object{PSR B1259-63},
was   seen  with a flux of $\sim$~1~mCrab in 1994 with
\textit{ASCA} \citep{1995ApJ...453..424K}, while it was about a
factor ten brighter during  recent \textit{XMM-Newton}
observations carried out in January 2004. The
\textit{ISGRI} data show that \object{1RXP J130159.6-635806} had an 18-60 keV
flux of  $\sim$13 mCrab on February 7 and it faded below the
sensitivity threshold by February 19 \citep{2004ATel..251....1C}.

The comparison of the two panels of  Fig.~\ref{fig:image} clearly
indicates the importance of imaging observations to properly identify
the origin of the hard X-ray  flux measured from this region.  Other sources were detected in the large field of view
of \textit{ISGRI} during these observations (e.g. Cen~X-3, GX~301--2,
1E~1145.1--6461, the results on which will be reported
elsewhere) and were found at their
known coordinates.  Hence we can be confident that the
effect seen in Fig.~\ref{fig:image} is due to the variability of
the two sources and not to some problem in the attitude
reconstruction.

To study the spectrum of \object{PSR B1259-63} we extracted fluxes in
different energy ranges from the mosaic images and used the December
2003 version of the instrumental response matrix.  The resulting 
spectrum is shown in Fig.~\ref{fig:spec} and appears
consistent with those measured by other experiments. Because of the
small number of points and large errors, it his hard to constrain a
power law photon index, $\Gamma$, with this data and a fit with XSPEC
in the 20--200~keV range gives $\Gamma = 1.3 \pm 0.5$ (90\% confidence levels).  Although
seemingly harder, this is within the errors of the \textit{OSSE} spectrum, measured in the
20--300~keV range by \citet{1995ApJ...447L.113G}, of $\Gamma = 1.8 \pm
0.6$.

Timing analysis of  \textit{ISGRI} data, using the event list
method \citep{ebisawa}, did not reveal any
pulsations in the 20--60 keV energy band above the 3$\sigma$ level (but note that  the high
background level at these energies implies that the limits on the
pulsed fraction from such a search are not very constraining).

We also analyzed data from the \textit{INTEGRAL} spectrometer
\textit{SPI} \citep{2003A&A...411L..63V},
obtaining only a marginally significant detection ($\sim3\sigma$)
in the 20--40 and 40--100 keV bands. The SPI points, plotted as upper limits in Fig.~\ref{fig:spec},  are  consistent with the
\textit{ISGRI} detection of \object{PSR B1259-63}.

Finally,  \object{PSR B1259-63} appears to be too faint to be significantly
detected in the 3--20 keV range with the \textit{JEM-X} X-ray
monitor aboard \textit{INTEGRAL} \citep{2003A&A...411L.231L}.

\subsection{RXTE}
\object{PSR B1259-63} was monitored with  \textit{RXTE} during a public ToO
programme (n.P90405). The data from the \textit{Proportional
Counter Array (PCA)} instrument were reduced and analysed, using
the {\emph{LHEASOFT V5.3}} software package, as described in
\citet{2003ApJ...595.1032R}. In addition, time intervals contaminated by high fluxes of energetic
electrons were removed.

\begin{figure}
 \centering
 \resizebox{\hsize}{!}{\includegraphics{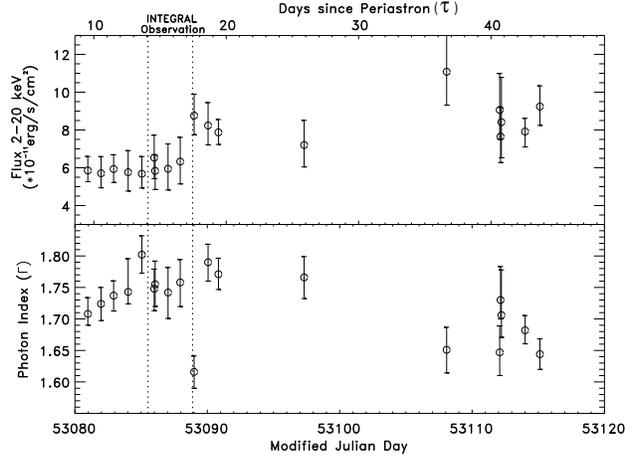}}
 \caption{Evolution of the flux (Top) and fitted power law spectral index (Bottom) of \object{PSR B1259-63}, measured in the 2-20 keV range with
   \textit{RXTE} after the 2004 periastron passage.  A
 constant galactic background component has been subtracted from the flux.  Four
 \textit{RXTE} pointings were simultaneous  with the \textit{INTEGRAL} ToO observation, indicated
 by the vertical dotted lines.  The top axis shows the time in days
 since periastron ($\tau = $ MJD 53071.43), which allows the \textit{INTEGRAL} observation
 ($\tau+14.1$ to $\tau+17.5$) to be compared with
 observations from the 1994 periastron epoch ($\tau = $ MJD 49361.15),
 discussed in Sec.~\ref{sec:intro}, with \textit{ASCA} (at $\tau +
 17.4$ to $\tau + 17.9$)
and \textit{OSSE} (at $\tau - 6$ to $\tau + 20$).}
 \label{fig:rxtelc}

 \end{figure}

The 2--20~keV \textit{PCA} light curve is shown
in the top panel of Fig.~\ref{fig:rxtelc}. Due to the source
faintness,  some contamination by the Galactic Ridge X-ray
emission in the  $\sim1^{\circ}\times1^{\circ}$ field of view
of the \textit{PCA} collimator is likely. To
estimate it we assume that the  Galactic Ridge contribution  at the
position of \object{PSR B1259-63} is the same as that of region R1 of
\citet{1998ApJ...505..134V}, corresponding to
 $\sim2.4\times10^{-11}$ erg~cm$^{-2}$~s$^{-1}$.

The lower panel of Fig.~\ref{fig:rxtelc} shows the photon index
obtained by fitting the  \textit{PCA} data  with an absorbed power law, with
$\rm{N}_{\rm{H}}$ frozen to $0.6\times 10^{22}$~cm$^{-2}$, plus an iron line at 6.5 keV \citep{1999ApJ...521..718H}. The
power law photon index was found to be $\sim$1.6--1.8 in all the
observations, which is consistent with earlier measurements.

Four points in the light curve were simultaneous with the
\textit{INTEGRAL} ToO observation and, since their fluxes and
fitted spectral indices are consistent with a constant value, it
was possible to make an average \textit{RXTE} spectrum. The Galactic background was subtracted before the
spectrum was fitted in XSPEC (v11.3). With $\rm{N}_{\rm{H}}$ frozen to 0.6$\times10^{22}$~cm$^{-2}$ \citep{1999ApJ...521..718H}, we obtain a harder photon index of
1.5, while a flattening is obvious at low energies
(see Fig.~\ref{fig:spec}). Allowing $\rm{N}_{\rm{H}}$ to vary, leads
to a photon index consistent with that of the individual spectra,
and previous findings although $\rm{N}_{\rm{H}}$ tends to much
higher values. This is interpreted as evidence of a slight
overestimate of the background, especially at low energies.
Restraining the fit to the 5-30 keV  energy range, with
$\rm{N}_{\rm{H}}$ frozen to its normal value, leads to
$\Gamma=1.71\pm0.08$ (90\%) and $\chi_\nu=1.07$ (48 dof).

The non detection of \object{1RXP J130159.6-635806} in the \textit{ISGRI}
images of March,  the relative steadiness of the RXTE flux, and
the relatively high absolute value of the power law photon index
suggest that the \textit{RXTE} flux is dominated by emission from
\object{PSR B1259-63}.  However, due to the lack of imaging of the \textit{PCA} instrument, we
cannot exclude contribution from  1RXP~J130159.6-635806 at other times
(e.g. the flux increase seen at MJD 53089).

\section{Discussion}

\label{sec:results}

\begin{figure}[t]
\centering
\resizebox{\hsize}{!}{\includegraphics{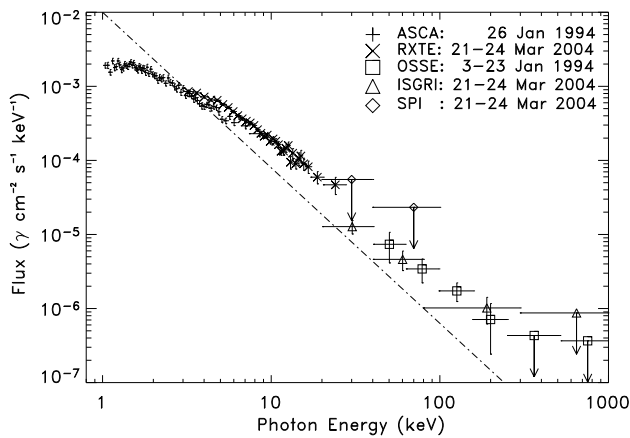}}
\caption{Spectral Energy Distribution for \object{PSR B1259-63} obtained
from several  instruments: Results from the \textit{INTEGRAL} instruments,
 \textit{ISGRI} and \textit{SPI}, are plotted with contemporaneous
 \textit{RXTE} observations for the March 2004 post-periastron disk
 passage (this work, see also Fig.~\ref{fig:rxtelc}).  \textit{ASCA} observations of 26 January
1994 are also shown \citep{1999ApJ...521..718H}, along with the
integrated \textit{OSSE} spectrum for the period 3--23 January
1994 \citep{1995ApJ...447L.113G}.  For
comaprison, the dot-dashed line corresponds to a Crab-like
spectrum with 1 mCrab brightness.} \label{fig:spec}
\end{figure}

The spectral energy distribution for \object{PSR B1259-63}, combining these
and previous soft gamma~/~hard X-ray measurements, is shown in
Fig.~\ref{fig:spec}.  The flux measured with \textit{ISGRI} corresponds to
$\sim$2~mCrab at 30~keV, and rises to $\sim$3~mCrab and
$\sim$6~mCrab at 60
and 190~keV respectively.  The average luminosity, given that the system is 1.5~kpc from
Earth \citep{1996MNRAS.279.1026J}, is estimated as $(8.1 \pm
1.6)\times10^{33}$~erg~s$^{-1}$ in the 20--80~keV band.  The \textit{ISGRI} spectrum is in  good agreement with the extrapolation of the
simultaneous \textit{RXTE/PCA} spectrum and is consistent with that
measured above 30 keV with \textit{OSSE} during the  1994
periastron passage. However, contrary to the  \textit{OSSE} data, the
\textit{ISGRI} observation has the great advantage of clearly
identifying the \object{PSR B1259-63}~/~\object{SS2883} system as the source
responsible for the observed hard X-ray emission.

Also  shown in Fig.~\ref{fig:spec} is a spectrum taken during the
1994 post-periastron disk passage  ($\tau +17.4$ days) with the
\textit{ASCA} X-ray telescope, which can be compared with the
\textit{INTEGRAL} observations ($\tau+14.1$ to $\tau+17.5$ days).  \citet{1999ApJ...521..718H} suggested a potential hardening of the
\textit{ASCA} spectrum, from a photon index of 1.96 at periastron
to 1.69 at $\tau +17.4$ days.  The \textit{RXTE} data,
shown in Fig.~\ref{fig:rxtelc}, also indicate a hardening of the
spectrum, which seems to be accompanied by a slight increase in
flux, at $\tau +17.5$ days.  Unfortunately, this occurs just after
the \textit{INTEGRAL} observation and contamination  by 1RXP~J130159.6-635806 can not be ruled out without the high resolution
\textit{ISGRI} images.

Several models have been put forward to explain the high energy emission
from \object{PSR B1259-63}.  Although all involve the interaction between the two stars
they differ in the invoked mechanisms for the emitted radiation, as the
properties of pulsar relativistic winds are a matter of debate.

\citet{1997ApJ...477..439T} suggest a
high Lorentz factor of the pulsar wind, $\gamma_{w} \sim 10^{6}$, with
X-rays from synchrotron emission of relativistic particles in the
pulsar wind.  In this case the non-pulsed radio emission
is explained by synchrotron emission from the electrons of the disk of
the Be star, accelerated at a shock wave which appears during
the passage of the neutron star through the disk \citep{1999ApJ...514L..39B}.
\citet{1999MNRAS.304..359C,2000Ap&SS.274..177C} suggest a more
moderate  $\gamma_{w}$ of 10-100 with the relativistic particles of
the pulsar wind being responisble for both the X-ray (inverse Compton
scattering of the 
pulsar wind on Be star soft photons) and non-pulsed radio emission
(synchrotron radiation).

For the source of the TeV emission, \citet{1999APh....10...31K} propose relativistic
electrons from the pulsar wind, while
\citet{2004ApJ...607..949K} favour interactions of protons accelerated
in a shock wave during the passage of the neutron star through
the Be-star disk.  As the hard X-ray and TeV emission may originate
from different populations of particles, TeV observations alone
may be unable to constrain $\gamma_{w}$ and the post-periastron data
reported here also leaves this question open.  A report on the full
\textit{HESS} observation of \object{PSR B1259-63} around periastron
is being produced (Aharonian, et.~al., in prep.).

Our observations underline the need for broad band simultaneous
coverage, with instruments able to resolve source confusion problems,
to investigate the high energy emission of this source in detail.  The combination of the post-periastron observations reported here,
with pre-periastron \textit{XMM-Newton} data, which show a sudden
increase of the X-ray intensity during the first disk crossing, may be
enough to rule out some models (Chernyakova, et.~al., in prep.).

Finally we note that the data reported here show that \textit{INTEGRAL} has
the capability of detecting \object{PSR B1259-63} with observing times
about a factor 10 shorter than \textit{OSSE}.  More intensive and
coordinated hard X-ray observations of the future periastron
passages will certainly provide useful constraints and lead to the
possible detection of flux and spectral variability on time scales
of days, which might help to discriminate among the different
models.

\begin{acknowledgements}
This paper is based on observations made with the ESA \textit{INTEGRAL} project
and also uses data obtained from the HEASARC Online Service, provided
by the NASA Goddard Space Flight Center.  Individual authors thank the
following bodies for financial support:  JR, the French Space Agency 
(CNES); SES, UK PPARC.
\end{acknowledgements}

\bibliographystyle{aa}
\bibliography{Gg061}

\end{document}